\documentclass[aps,prb,twocolumn,superscriptaddress]{revtex4-2}

\usepackage{bm}
\usepackage{graphicx}
\usepackage{subfigure}
\usepackage{amssymb}
\usepackage{amsmath}
\usepackage{verbatim}
\usepackage{color}

\begin{document}

\title{Spin dynamics of the quantum dipolar magnet Yb$_{3}$Ga$_{5}$O$_{12}$ in an external field} 

\author{E. Lhotel}
\affiliation{Institut N\'eel, CNRS \& Univ. Grenoble Alpes, 38000 Grenoble, France}
\author{L. Mangin-Thro}
\affiliation{Institut Laue Langevin, 38042 Grenoble, France}
\author{E. Ressouche}
\affiliation{Univ. Grenoble Alpes, CEA, IRIG, MEM, MDN, 38000 Grenoble, France}
\author{P. Steffens}
\affiliation{Institut Laue Langevin, 38042 Grenoble, France}
\author{E. Bichaud }
\affiliation{Univ. Grenoble Alpes, Grenoble INP, CEA, IRIG, PHELIQS, 38000 Grenoble, France}
\author{G. Knebel}
\affiliation{Univ. Grenoble Alpes, Grenoble INP, CEA, IRIG, PHELIQS, 38000 Grenoble, France}
\author{J.-P. Brison}
\affiliation{Univ. Grenoble Alpes, Grenoble INP, CEA, IRIG, PHELIQS, 38000 Grenoble, France}
\author{C. Marin}
\affiliation{Univ. Grenoble Alpes, Grenoble INP, CEA, IRIG, PHELIQS, 38000 Grenoble, France}
\author{S. Raymond}
\email{raymond@ill.fr}
\affiliation{Univ. Grenoble Alpes, CEA, IRIG, MEM, MDN, 38000 Grenoble, France}
\author{M. E. Zhitomirsky} 
\email{mike.zhitomirsky@cea.fr}
\affiliation{Univ. Grenoble Alpes, Grenoble INP, CEA, IRIG, PHELIQS, 38000 Grenoble, France}

\date{\today}

\begin{abstract}
We investigate ytterbium gallium garnet Yb$_{3}$Ga$_{5}$O$_{12}$ in the paramagnetic phase 
above the supposed magnetic transition at $T_{\lambda} \approx 54$~mK. Our study combines susceptibility 
and specific heat  measurements with neutron scattering experiments and theoretical calculations.
Below 500 mK, the elastic neutron response is strongly peaked in the momentum space. Along with that 
the inelastic spectrum develops flat excitation modes.  In magnetic field, the lowest energy branch follows 
a Zeeman shift in accordance with the field-dependent specific heat data. 
An intermediate state with spin canting away from the field direction is evidenced in small magnetic fields.
In the field of 2 T, the total magnetization almost saturates and
the measured excitation spectrum is well reproduced by the spin-wave calculations 
taking into account solely the dipole-dipole interactions. 
The small positive Curie-Weiss temperature derived from the susceptibility measurements
is also accounted for by the dipole spin model. Altogether, our results suggest that 
Yb$_{3}$Ga$_{5}$O$_{12}$ is a quantum dipolar  magnet.
\end{abstract}

\maketitle

\section{Introduction}

Geometrical frustration typically suppresses conventional magnetic ordering and 
may stabilize exotic phases with unconventional spin excitations \cite{Balents10,Broholm20}. 
Apart from fundamental interest, the delayed magnetic ordering in conjunction with 
a large unfrozen entropy makes frustrated materials to be good candidates for the low-temperature 
magnetic cooling \cite{Zhito03}.
The developing space applications and increasing costs of helium fuel a continuing search for 
new refrigerant materials for the adiabatic demagnetization refrigeration (ADR) in 
the 0.1--4 K temperature range, which can outperform the presently used paramagnetic salts \cite{Wikus}.
In the past few years there have been experimental reports of the enhanced magnetocaloric effect observed
for pyrochlores \cite{Sosin05,Orendac07,Wolf16}, intermetallics \cite{Pakhira17,Jang15,Tokiwa16},
and various Yb compounds \cite{Jang15,Tokiwa16,Paixao,Tokiwa21}.

In this work, we investigate ytterbium gallium garnet Yb$_{3}$Ga$_{5}$O$_{12}$, which exhibits
an enhanced magnetocaloric effect below 2~K \cite{Paixao}. There is also a reborn interest 
in the rare-earth gallium garnets $R_{3}$Ga$_{5}$O$_{12}$ in general. The recent experimental studies
focus on  $R = \rm Er$ \cite{Cai}, $R = \rm Tb$ \cite{Wav,Petit}, and $R = \rm Dy$ \cite{Kibalin} compounds, 
whereas gadolinium gallium garnet  was extensively studied in the past two decades, see, for example, 
\cite{Paddison} and references therein.

In the cubic garnet crystals, space group $Ia\bar{3}d$, the rare-earth ions form 
two inter-penetrating lattices of corner-sharing triangles often called hyperkagome lattices.
The hyperkagome lattice strongly frustrates the nearest-neighbor antiferromagnetic
exchange  keeping a large degree of frustration in the presence of anisotropy  and dipolar interactions
as well. Positions of the crystal field levels of Yb$^{3+}$ ions 
were measured by Buchanan {\it et al.}\ for several ytterbium garnets using infrared spectroscopy \cite{Buchanan67}.
The lowest $J=7/2$ multiplet is split into four Kramers doublets by the local  orthorhombic crystal field.
In Yb$_{3}$Ga$_{5}$O$_{12}$, the distance between the lowest
doublet and the next excited level is about 67~meV. Hence, at temperatures below 10~K, one can safely neglect
the excited levels and interpret magnetic properties in terms of the effective $S=1/2$ degrees of
freedom corresponding to the ground state doublet. The magnetic entropy variation $\Delta S_m =R\ln 2$
determined in the specific heat measurements between 44~mK and 4~K \cite{Filippi80} is fully consistent with 
this conclusion. The $g$-tensor of the ground-state doublet is only weakly anisotropic 
 as follows from EPR measurements  \cite{Carson60} and crystal-field calculations 
\cite{Pearson67}. 

Below 4~K, the magnetic susceptibility of Yb$_{3}$Ga$_{5}$O$_{12}$  follows the Curie-Weiss law  with 
a small Curie-Weiss temperature  $\theta_{p}=45$~mK  \cite{Filippi80}.  In addition, the specific heat exhibits 
a lambda-peak anomaly at $T_\lambda = 54$~mK preceded by a broad maximum at 200 mK. 
These features were tentatively associated with a long-range order and short-range spin correlations, 
respectively \cite{Filippi80}. However, M\"ossbauer spectroscopy measurements did not detect any hyperfine splitting 
associated with a possible magnetic ordering below $T_\lambda$ \cite{Hodges}. 
$\mu$SR experiments did not observe signs of a long-range order as well pointing instead to the reinforcement 
of the dynamical correlations  below $T_\lambda$  \cite{Dalmas03}. 
The diffuse magnetic scattering in Yb$_{3}$Ga$_{5}$O$_{12}$ has been recently measured in
neutron experiments \cite{Sandberg20} and interpreted in terms of the exotic director state associated with  ten-site loops
 by analogy with a similar suggestion for Gd$_{3}$Ga$_{5}$O$_{12}$ \cite{Paddison}. 
The experimental studies in Ref. \cite{Sandberg20} have been complemented by simulations of the exchange-dipolar
Ising  model.
Thus, despite several decades of research, the basic magnetic  properties of ytterbium gallium garnet remain poorly
understood both in terms of possible magnetic states  \cite{Capel} as well as underlying
microscopic interactions.
In the present paper, we combine results of bulk measurements of single crystals of Yb$_{3}$Ga$_{5}$O$_{12}$ 
with neutron scattering experiments in an applied magnetic field and theoretical 
calculations to get a further insight into the microscopic properties of this interesting frustrated material.
The combination of complementary techniques allows us to quantitatively 
conclude on a dominant role of the dipolar interactions both for the static and dynamic
properties of this interesting spin-1/2 hyperkagome material.

\section{Experimental details}

\subsection{Sample preparation}

Single crystals of Yb$_3$Ga$_5$O$_{12}$ were grown by the floating zone technique using 
a commercial optical furnace (Crystal System Inc.)\ equipped with 1 kW halogen lamps. The feed rods were 
prepared from high-purity oxides Yb$_2$O$_3$ (99.99\%) and Ga$_2$O$_3$ (99.99\%) mixed and heated  
up to 1150$^{\circ}$C. After sintering, the rods were pressed isostatically at 1500 bar and sintered again 
at 1400$^{\circ}$C in air. Crystal growth conditions were optimized under argon pressure (4.5 bar) 
with a gas flow of 0.5~L/min at the growth rate of 2 mm/h. The feed and crystallized rods have been rotated  
in opposite directions at the rate of 30 rounds per minute. Crystals of Yb$_{3}$Ga$_{5}$O$_{12}$ prepared with 
this procedure grow in a direction very close to a four-fold axis of the cubic $Ia{\bar 3}d$  phase.
We observe that rods are white transparent when grown under air or oxygen (oxidizing) 
conditions or slightly pink under argon (reducing) atmosphere. No foreign phase was detected in the $X$-ray 
powder diffraction experiments. The checks of crystal quality and the crystallographic orientation were done 
using the Laue diffraction technique. It should be mentioned that the crystal color can be reversibly changed 
from white to pink according to the post growth heat treatment in a specific atmosphere. No significant 
modifications of the basic physical properties (magnetization, specific heat) were observed between both states.

Growth of single crystals was complemented by preparation of a high-density ceramic of
Yb$_{3}$Ga$_{5}$O$_{12}$ suitable for a direct use in an ADR prototype \cite{Paixao}.
Ceramic samples were made from the same starting powders of Yb$_{2}$O$_{3}$ and Ga$_{2}$O$_{3}$. 
An extensive investigation was carried out using a planetary crusher to perform fine grinding followed by
laser granulometry to measure a powder-diameter distribution, and high-temperature 
dilatometry to determine optimal temperatures for synthesis and sintering. 
After all, the starting materials were first treated at 1050$^{\circ}$C before being pressed isostatically 
under 2500 bar and, then, sintered at 1700$^{\circ}$C. The ceramic samples were checked 
by $X$-ray powder diffraction, showing the same cubic $Ia\bar{3}d$ phase.
No traces of foreign phases have been detected. 
Following this optimized route, the density of ceramic samples reaches 96(1)\%\ of the bulk crystal density showing 
no open porosity as evidenced by scanning electronic microscopy.

\subsection{Methods}

The specific heat was measured on a commercial Quantum Design Physical Property Measurement System 
with the $^{3}$He insert using the heat-pulse semi-adiabatic technique. Two types of samples were investigated: 
 a 0.87 mg ceramic sample in fields up to 8 T and a 17.25 mg single crystal (white color) with the field applied along 
 the [001] direction in fields  up to 3 T.

Magnetization was measured using the home-made low temperature SQUID magnetometer developed at the Institut N\'eel, 
which operates in a dilution fridge \cite{Paulsen}. A  single crystal of Yb$_3$Ga$_5$O$_{12}$ shaped as a
parallelepiped of size $0.93 \times 0.95 \times 4.37$~mm$^3$ has  been glued on a copper sample holder with 
GE varnish to ensure a good thermalization. Magnetic field was applied along the long side of the sample, 
which corresponds to the [001] direction. The corresponding demagnetization factor was estimated to 
1.19 (cgs units) \cite{Aharoni98}.  All the data shown below are corrected from demagnetizing effects. 

The neutron scattering experiments were performed at the high-flux reactor of the Institut Laue-Langevin, 
Grenoble. Diffraction measurements were done on the two-axis thermal-neutron diffractometer D23 
equipped with a lifting detector. A copper monochromator provides an unpolarized beam with the wavelength 
of 1.276~\AA. The diffuse scattering was measured on the D7 beam line for the polarization along the vertical $z$-axis
and  the wavelength of the incident beam being 4.8~{\AA}.
A four-to-one time ratio was used between the spin-flip and non spin-flip channel measurements. 
The data were collected in a 120$^{\circ}$ range, with steps of 1$^{\circ}$ for two positions of the detector banks, 
resulting in 0.5$^{\circ}$ between each measured position. Standard vanadium, quartz and empty corrections have been 
performed for taking into account the detector efficiency, the polarization efficiency and the background scattering \cite{Stewart}. 
The flipping ratio measured in the direct beam was in the range 20-29 during all the experiment. However the flipping ratio was larger 
at higher temperatures (20-27 at 50 mK and 25-29 at 5 K).

The inelastic neutron-scattering (INS) experiments were carried out on the three-axis spectrometer ThALES. 
The instrument was used in the classical single detector mode and set up in the W configuration. 
The incident beam was provided by the Si monochromator with a velocity selector in place. 
A Be filter was put in the scattered beam and a radial collimator was placed between the graphite analyzer and the detector. 
Most of the data were taken at fixed $k_f =1.2$~\AA$^{-1}$ (corresponding to 5.2~\AA).

The white-color crystal of Yb$_{3}$Ga$_{5}$O$_{12}$ used in the neutron experiments has a cylindrical shape 
(diameter 8 mm, length 27 mm) with the long axis parallel to the [001] direction. The sample
was embedded in a copper foil glued with hydrogen-free fluorinated Fomblin oil, tightened with Teflon tape 
 to ensure good thermalization, and mounted in a dilution insert. 
For D23, a smaller part cut from this rod was used. The scattering plane was selected to be (1 0 0)-(0 1 0). 
For D23 and ThALES, the sample was inserted inside 6 T and 2.5 T vertical magnets respectively, so that the field was 
applied along the [001] direction. In our experiments, the lowest temperature indicated on the dilution fridge thermometer was 50 mK. 
Still, due to the distance between the sample and the dilution mixing chamber as well as to a heating effect 
from the neutron beam, we assume that our measurements were performed above 54 mK, 
which is the expected transition temperature for Yb$_{3}$Ga$_{5}$O$_{12}$ \cite{Filippi80}. 
For the sake of simplicity, we nevertheless use the thermometer temperatures in the description of our results.
In the following, the scattering vectors are expressed as $\textbf{Q}=(Q_h, Q_k, Q_l)$ and reciprocal space directions 
are labelled by $(h\ k\ l)$.

\subsection{Magnetic characterization}
\label{mag_charac}

\begin{figure}[ht]
\centering
\includegraphics[width = 0.95\columnwidth]{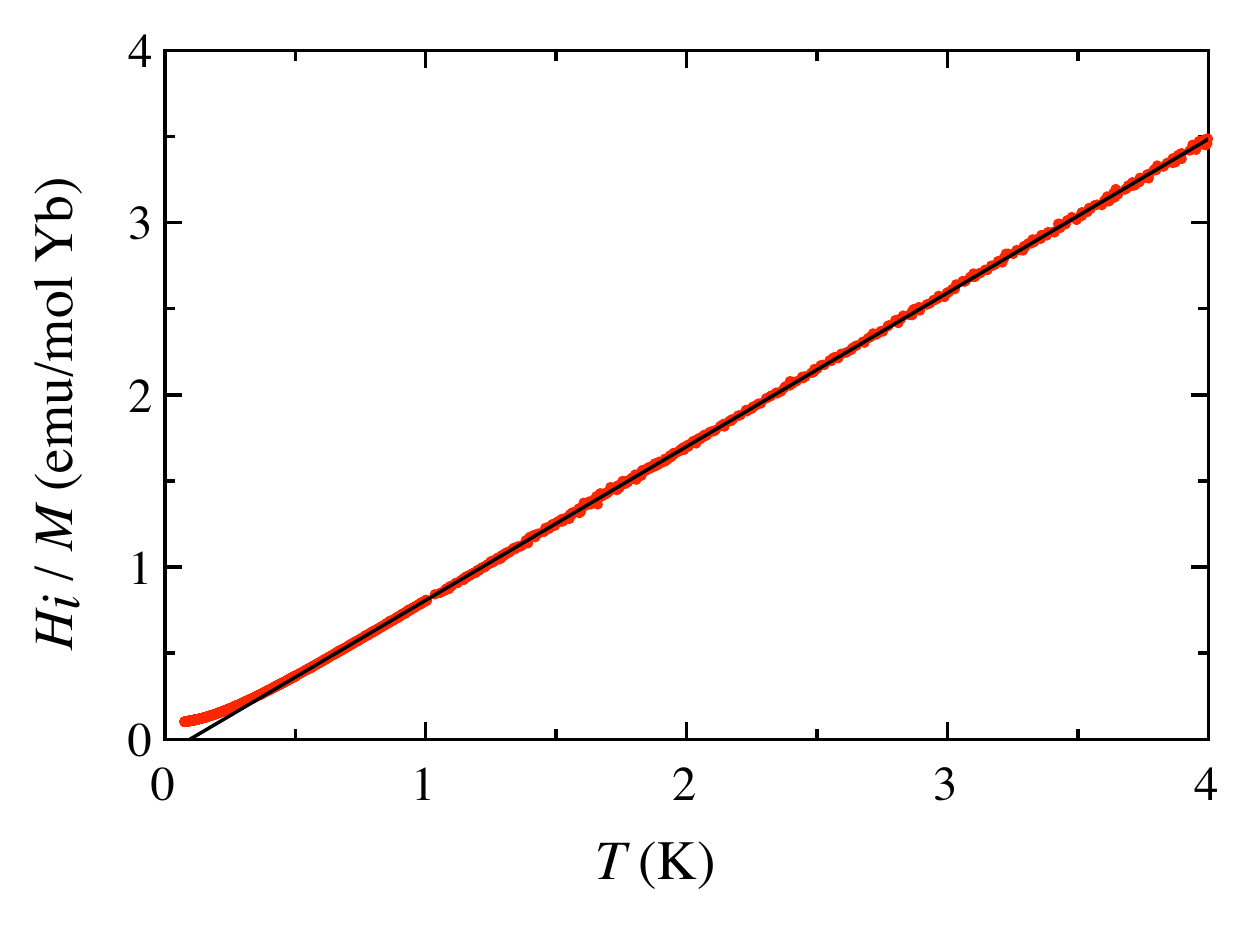}
\caption{Inverse magnetic susceptibility $\chi^{-1} = H_i/M$ as a function of temperature 
measured in the applied field $\mu_0 H_e = 0.01$~T oriented parallel to the [001] axis. 
The solid line is a fit to the Curie-Weiss law.}
\label{chi}
\end{figure}

The static magnetic susceptibility was measured in a weak applied field $\mu_0 H_e =0.01$~T
to insure a linear response in the whole range of temperatures down to 70~mK. The susceptibility
was calculated by dividing a uniform magnetization to an internal field $H_i$, which was obtained
from $H_e$ by applying the demagnetization correction at each temperature. 
The inverse susceptibility can be fitted in the 1--4.2~K range to a Curie-Weiss law 
$\chi^{-1}=(T-\theta_p)/C$, where $\theta_p$ is the Curie-Weiss temperature and $C$ is the Curie constant (see Fig. \ref{chi}). 
The fitting yields $\theta_p= 97$ mK and $C=1.12$ emu/mol Yb. The Curie constant corresponds to an effective moment 
of $3~\mu_{\rm B}$ for an Yb$^{3+}$ ion in agreement with the early measurements by Fillipi {\it et al.} \cite{Filippi80}.
Our value for the Curie-Weiss temperature is somewhat larger than their estimate $\theta_p\simeq 45$~mK,
though both agree on its ferromagnetic sign. In contrast, a negative (antiferromagnetic) $\theta_p$ has been recently 
reported in Ref. \onlinecite{Sandberg20}. Note, that accounting for the Van Vleck contribution, as was done in Ref. \onlinecite{Filippi80}, 
only slightly affects the values of $C$ and $\theta_p$. 

The saturated magnetization measured at 120 mK and 8 T is found to be 1.76 $\mu_{\rm B}$, corresponding to an effective $g=3.52$, 
in agreement with the previous reports and the crystal field calculations \cite{Filippi80, Carson60, Pearson67}.

\section{Diffuse scattering}

\begin{figure}
\begin{center}
\includegraphics[width = 0.95\columnwidth]{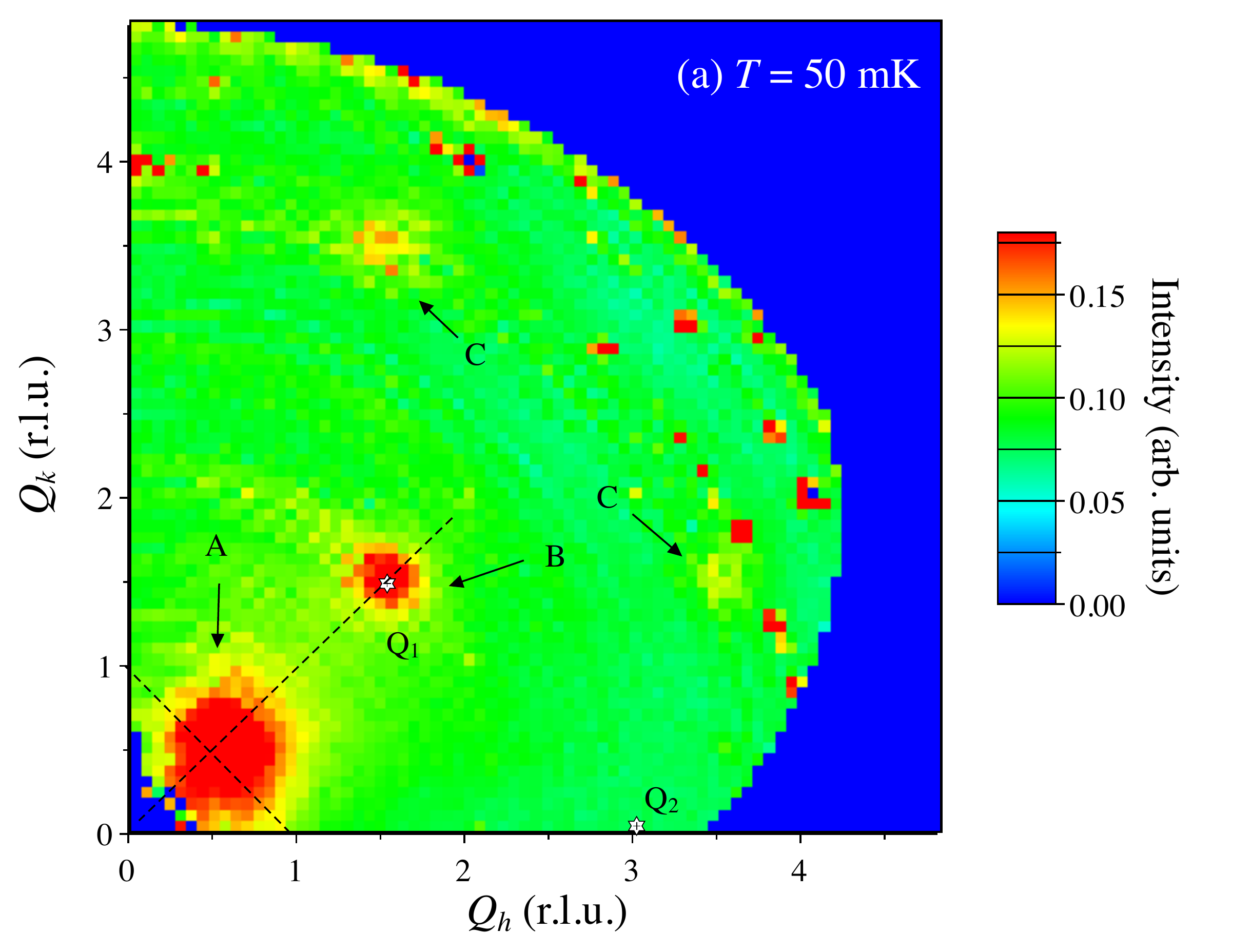} \\[4mm]
\includegraphics[width = 0.99\columnwidth]{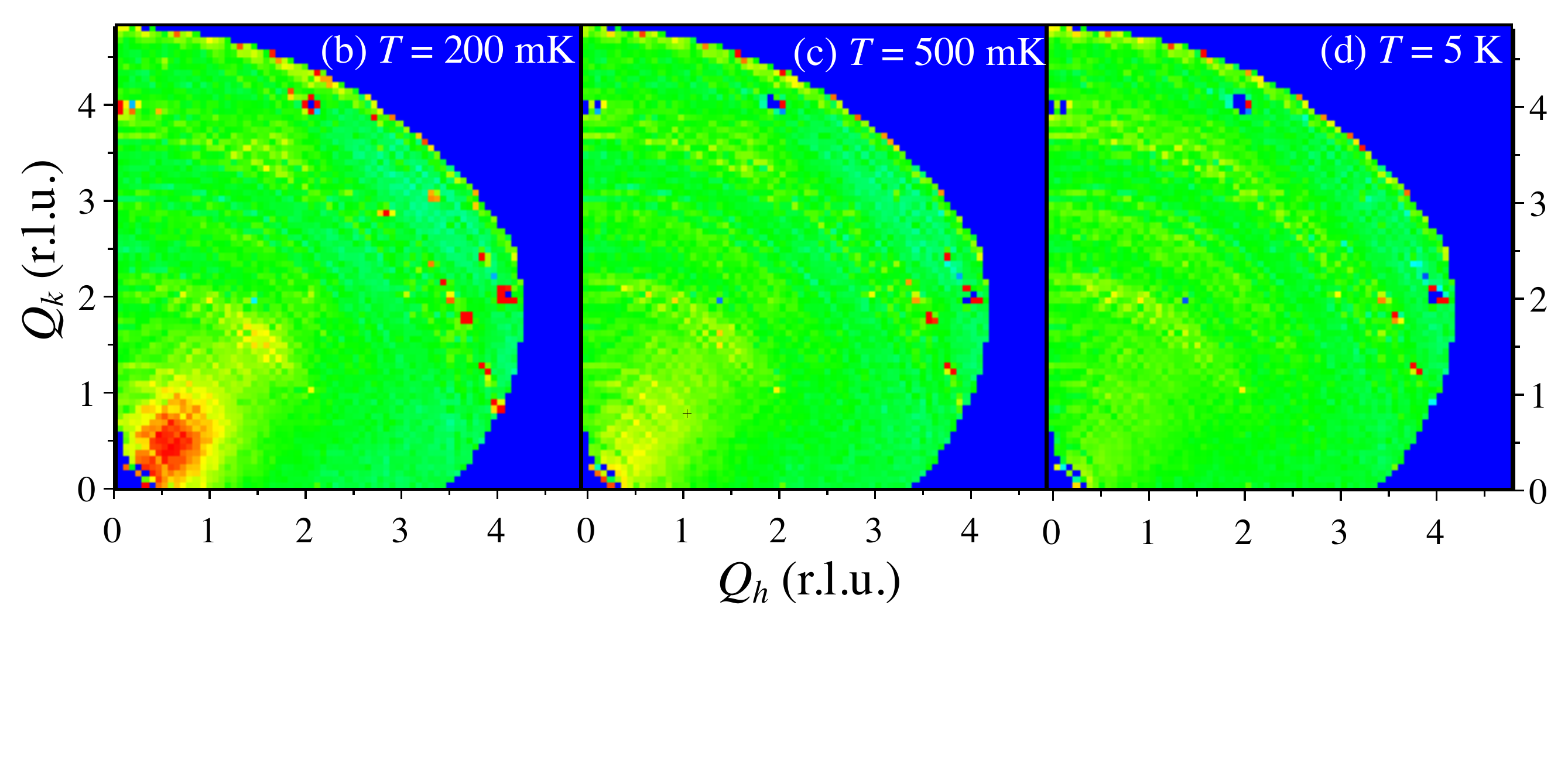}        
\end{center}
\vspace{-10mm}
\caption{Color-coded intensity plot of the magnetic scattering measured on D7 at (a) 50 mK, 
(b) 200 mK, (c) 500 mK, and (d)~5~K. The labels A, B and C in panel (a) mark the diffuse 
scattering features described in the text and the stars indicate the location of the momentum 
transfer ${\bf Q}_1 =(3/2, 3/2, 0)$ and ${\bf Q}_2=(3, 0, 0)$ investigated by INS. 
The dashed lines indicate the longitudinal and transverse cuts shown in Fig. \ref{Elast2}.}
\label{Elast1}
\end{figure}

The magnetic scattering measured on D7 in the spin flip channel with polarization along the $c$-axis is shown 
in Fig. \ref{Elast1} for four different temperatures 50 mK, 200 mK, 500 mK and 5 K.
The measured magnetic signal corresponds to the scattering obtained for an energy integration of the scattering function 
in the range [-$k_{B}T$, $E_i$], with $E_i$=3.5 meV.

Magnetic correlations are observed along the (1 1 0) direction and are peaked at $Q \approx 0.36$ {\AA}$^{-1}$ (feature A) 
and $Q \approx 1.1$ {\AA}$^{-1}$ (features B). They are washed out progressively with temperature and a $\bf{Q}$-independent 
weaker signal is observed at 5 K. In addition a small signal with the same characteristics in temperature is observed for 
$Q \approx 1.95$ {\AA}$^{-1}$ (in the directions (3 7 0) and (7 3 0); features~C).
The observation of Bragg peaks at high momentum transfer is ascribed to the inaccuracy of the polarization correction 
which happens for these nuclear Bragg peaks that have a tremendous intensity in the non spin flip channel.

\begin{figure}
 \includegraphics[width = 0.85\columnwidth]{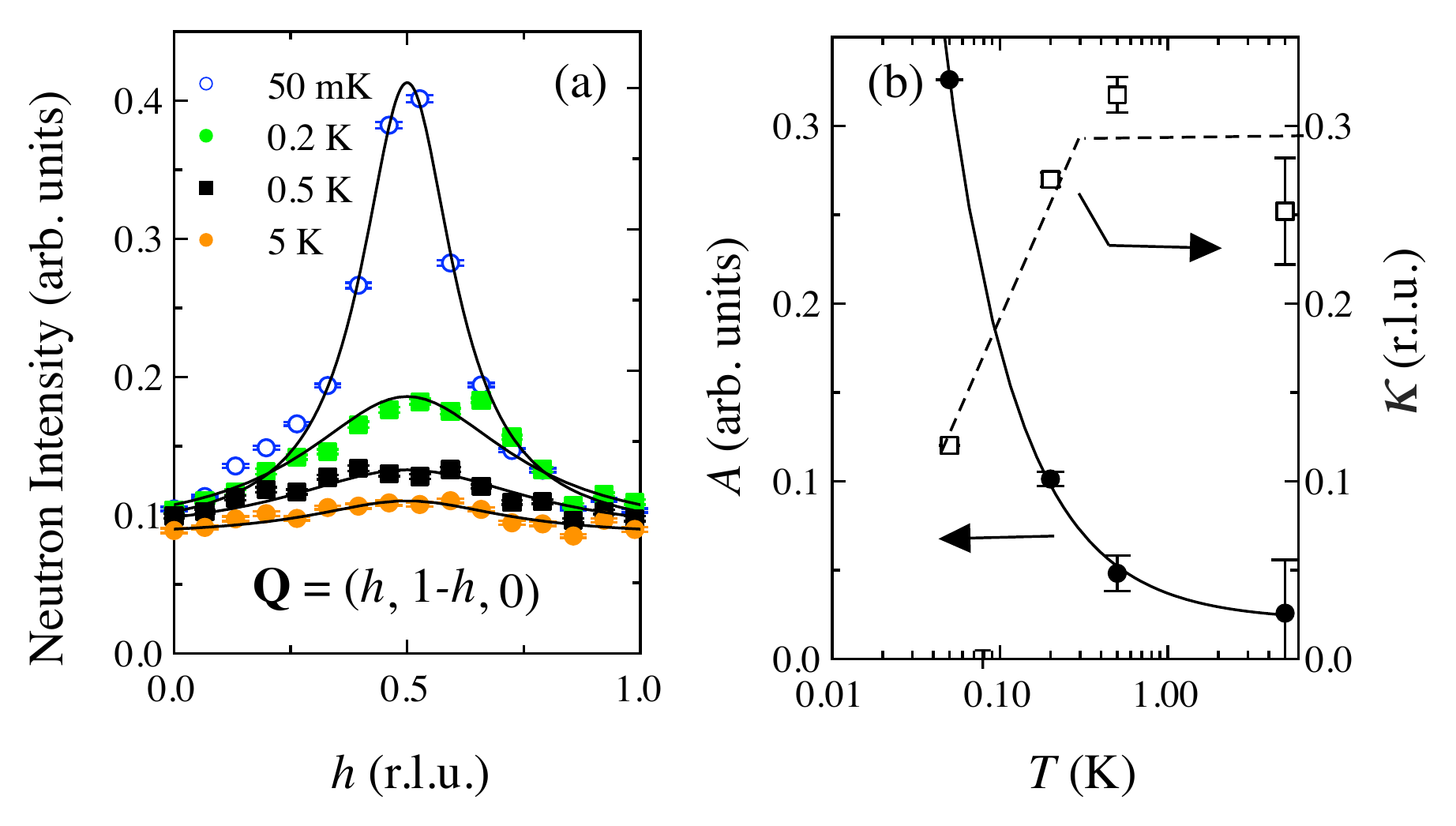}\\[2mm]
 \includegraphics[width = 0.85\columnwidth]{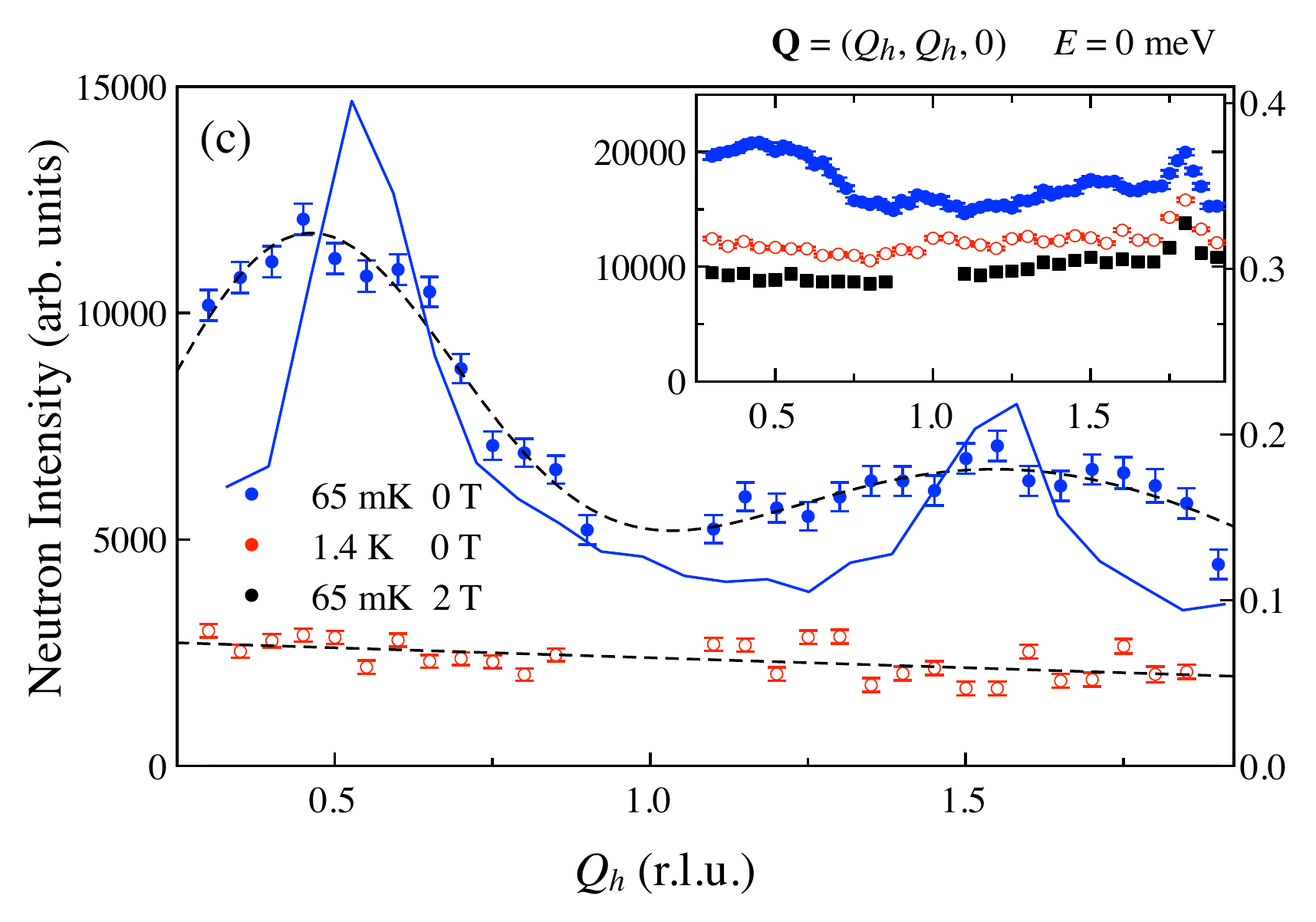}
\caption{(a) Momentum cut across feature A along the (1 -1 0) direction for several temperatures,
lines show the Lorentzian fits. 
(b) Temperature variation of the amplitude, $A$, and HWHM, $\kappa$, of the peaks shown in (a). 
Lines are guides to the eye. (c) Constant energy spectra performed on ThALES along ${\bf Q}=(Q_h$, $Q_h$, 0) 
at $E=0$~meV in zero field at $T=65$~mK and 1.4 K, subtracted from the 65 mK data for $\mu_0H=2$ T. 
The same data without subtraction are shown in the inset (in the 2 T data, the missing points in 
the figure correspond to the (1,~1,~0) magnetic Bragg peak induced under field). 
Dashed lines are guides to the eyes. In the main panel, the full line corresponds to the same longitudinal 
cut obtained from the D7 data at 50 mK. }
\label{Elast2}
\end{figure}

Figure~\ref{Elast2}(a) shows the peak obtained by performing a cut of the feature A in the direction 
(1 -1 0) for the four measured temperatures. The peaks are fitted by Lorentzian lineshapes and 
the resulting temperature dependence of the amplitude~$A$, and Half Width at Half Maximum (HWHM) $\kappa$, 
are shown in Fig. \ref{Elast2}(b). The signal clearly rises up upon cooling, and the correlation length increases 
below 200 mK. At 50 mK, the obtained HWHM is 0.12 r.l.u. (along (1 -1 0)), which corresponds to 
a correlation length, $\xi \propto 1/\kappa$, of about 11 {\AA}, {\it i.e.} to a distance between several 
Yb atoms in the cell (the minimum Yb-Yb distance being 3.74~\AA). 


The elastic scattering was measured on ThALES with an unpolarized neutron setup. 
Here, the measurement is made for zero energy transfer, $E=0$, and the integration in energy is performed over the resolution function, which Full Width at Half Maximum (FWHM) equals 0.039 meV as measured on the incoherent signal. 
The obtained scans along ${\bf Q}=(Q_h,\ Q_h,\ 0)$ at 65 mK and 1.4 K are shown in Fig. \ref{Elast2}(c). 

A modulation is observed at 65~mK, which is suppressed at 1.4 K. 
The observed signal at $E=0$ on ThALES corresponds to the features A and B highlighted on the D7 data as demonstrated by the similar peak positions in the figure. 
This indicates that the diffuse response is essentially dominated by true elastic or very low energy processes within the neutron time-scale. 
It is to be noted that the sample base temperature reached on ThALES is undoubtly higher than the one reached on D7 given the very high flux of cold neutron on the former instrument and this could explain the difference of lineshapes obtained owing to the strong temperature dependence of the signal below 200 mK.

\section{Zero field excitation spectrum}

The spin dynamics has been investigated along the high-symmetry lines and points of the Brillouin zone. 
Emphasis was put on specific wave-vectors at which the magnetic diffuse scattering described in the previous section 
was found to have its maximum or minimum intensity.
 
\begin{figure}
\centering
\includegraphics[width = 0.85\columnwidth]{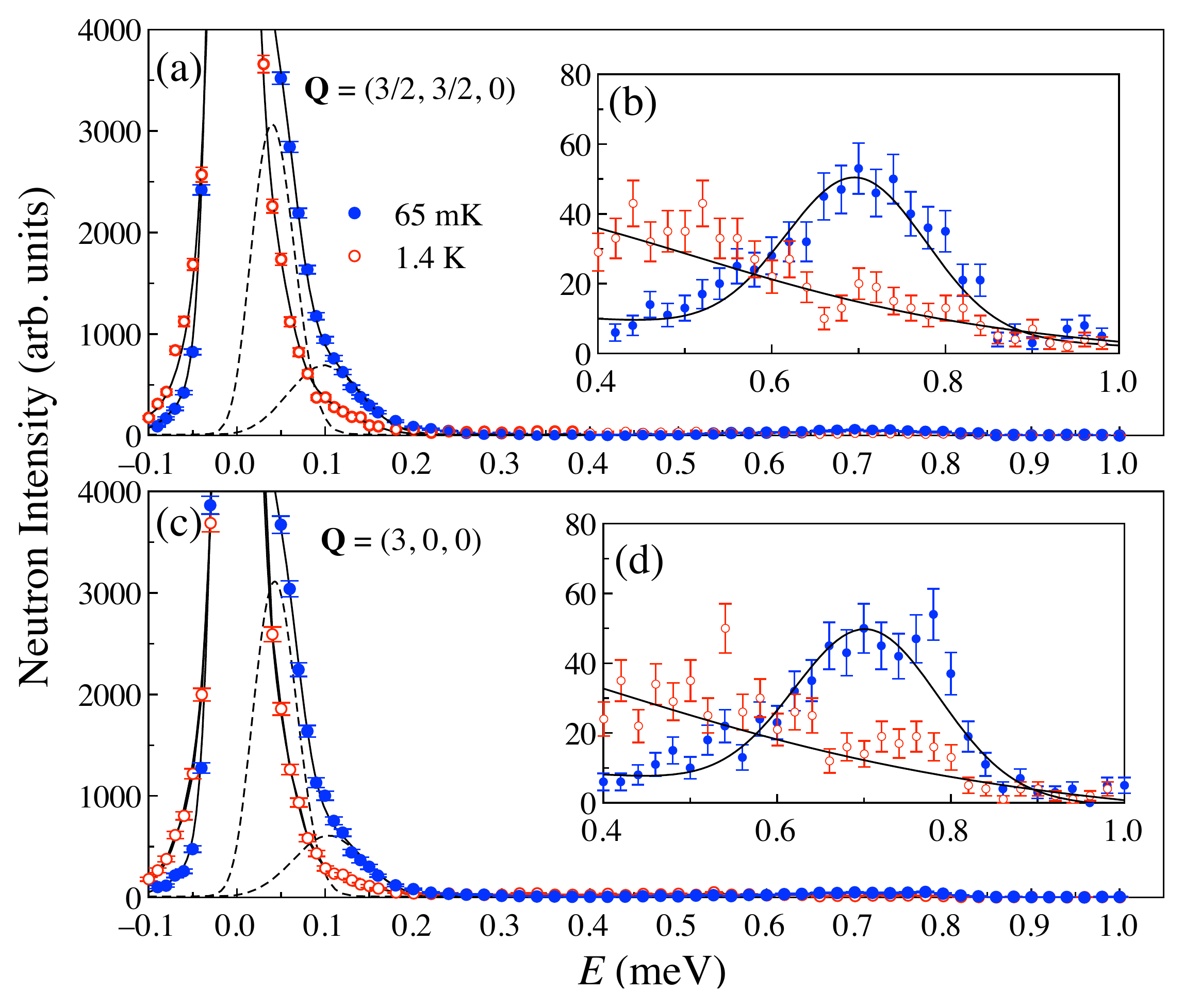}
\caption{Energy spectra at  $T= 65$~mK and 1.4 K in zero field: (a) and (b) for the scattering vector ${\bf Q}=(3/2,\ 3/2,\ 0)$ 
and (c) and (d) for ${\bf Q}=(3,\ 0,\ 0)$. Lines are fits as explained in the text.}
\label{INS1}
\end{figure}

Representative low energy spectra are shown in Fig.~\ref{INS1} for ${\bf Q}=(3/2,3/2,0)$, panels (a) and (b), and 
${\bf Q}=(3,0,0)$, panels (c) and (d), at $T= 65$~mK and 1.4 K. Positions of these two vectors are indicated by ${\bf Q}_1$ and
${\bf Q}_2$ in Fig.~\ref{Elast1}. At the lowest temperature, aside from the elastic line, the spectra are composed of several inelastic modes: a sharp mode at $E_0$, around 0.05 meV, a broader one at $E_1$, around 0.1 meV whose maximum intensity is barely resolved [panels (a) and (c)] and a very broad 
one at $E_2$, around 0.7 meV [panels (b) and (d)]. The constant $\textbf{Q}$-scans are phenomenologically described at low energy, 
panels (a) and (b), by a Gaussian lineshape, where both Stokes and anti-Stokes peaks together with the temperature factor are taken into account. 
A constant background and an elastic line with a fixed profile are also included (See Appendix \ref{appendix}). 

The fits shown in Fig.~\ref{INS1} are obtained with $E_0=0.045(6)$ meV and $E_1=0.100(6)$ meV and fixed FWHM of 0.059 meV and 0.094 meV respectively, which are about 2 and 3 times the incoherent width. The higher energy part 
is conveniently described by a broad Gaussian peak centered at $E_2=0.70(5)$ meV with a FWHM of 0.190(5) meV on a slopping background (See panels (b) and (d) of Fig. \ref{INS1}). 
At 1.4 K, the overall intensity is reduced in both the elastic and inelastic channels and a broad quasielastic-like signal replaces the well defined excitations. 
It is thus impossible to describe in a unique manner the lowest energy modes. The fits shown in Fig.~\ref{INS1} are indicative only and are obtained keeping the peak positions $E_0$ and $E_1$ to their low temperature values, while the peak widths and intensities are free. On the other hand, the spectral weight of the $E_2$ mode clearly moves to lower energies.

\begin{figure}
\centering
\includegraphics[width = 0.85\columnwidth]{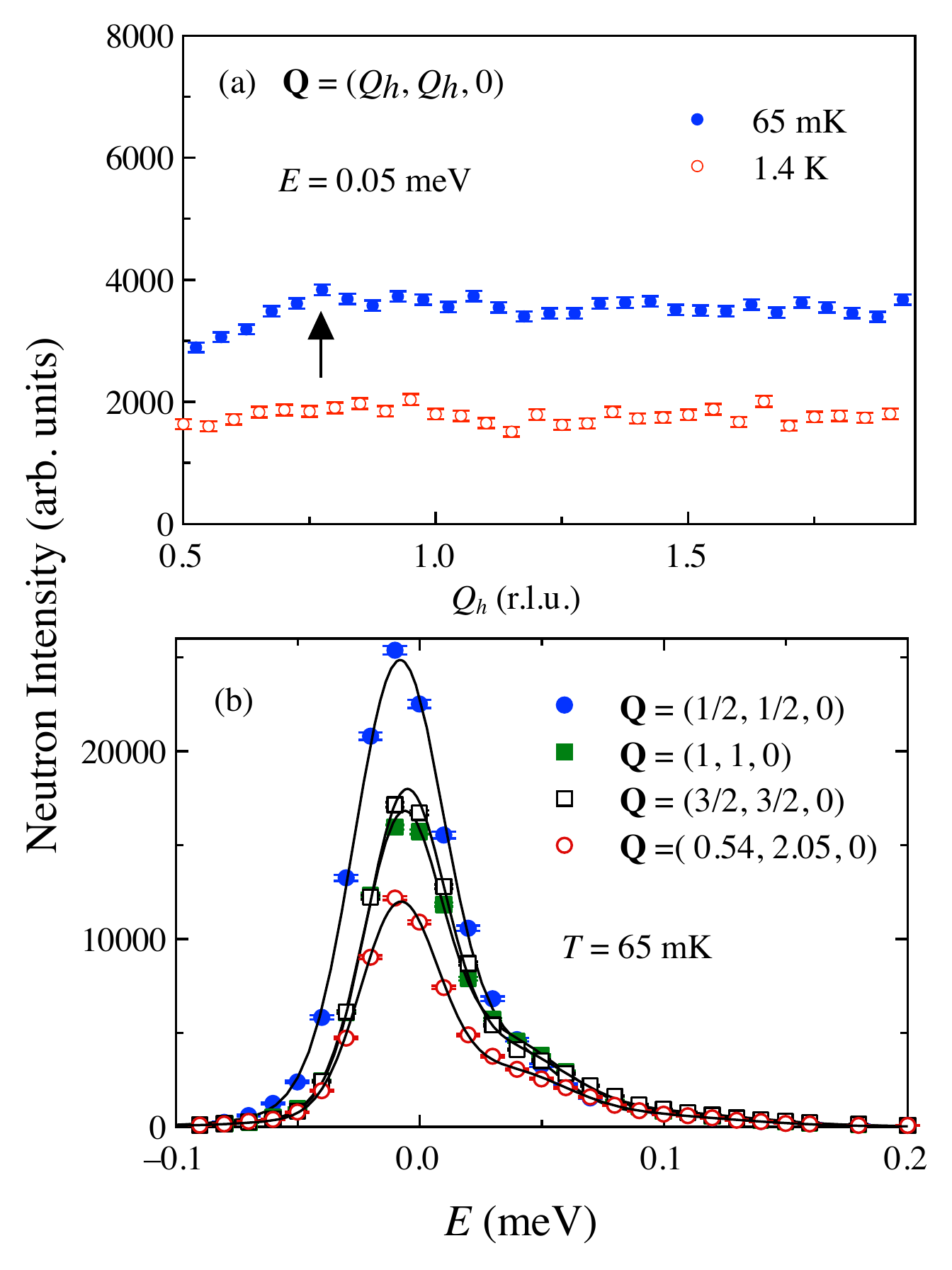}
\caption{(a) Constant energy scans performed along ${\bf Q}=(Q_h,\ Q_h,\ 0)$ for the energy transfer $E= 0.05$~meV at $T=  65$~mK and 1.4 K. 
(b) Energy spectra at ${\bf Q}=(Q_h,\ Q_h,\ 0)$ for $Q_h=1/2$, 1 and $3/2$ and at ${\bf Q}=(0.54,\ 2.05,\ 0)$ at 65~mK. Lines are fits performed 
using the same procedure as described for Fig.~\ref{INS1}. }
\label{INS2}
\end{figure}

The inelastic modes are essentially dispersionless and a representative constant energy scan performed along ${\bf Q}=(Q_h,\ Q_h,\ 0)$ for an energy transfer of 0.05 meV is shown in the panel (a) of Fig. \ref{INS2}. A broad intensity modulation with $\textbf{Q}$ is observed at low temperature, with a kink at $Q_h \approx 0.75$ r.l.u corresponding to a scattering vector magnitude of 0.55 {\AA}$^{-1}$ (arrow in Fig. \ref{INS2}(a)). The dispersionless nature of the excitations is further confirmed by several constant $\textbf{Q}$ scans performed along this line for $Q_h=1/2$, 1 and $3/2$ shown in Fig. \ref{INS2}(b) together with a scan performed at an arbitrary position $\textbf{Q}=(0.54,\ 2.05,\ 0)$ having the same $\textbf{Q}$ modulus as $\textbf{Q}=(3/2,\ 3/2,\ 0)$. Here again, only an intensity modulation is present, while no clear dispersion in energy could be detected.
In the remaining part of the paper, we focus  on the response at the scattering vectors $\textbf{Q}=(3/2, 3/2, 0)$ and $\textbf{Q}=(3, 0, 0)$.

\section{Magnetic response}
\label{mag_exp}


\begin{figure}[b!]
\centering
\includegraphics[width = 0.85\columnwidth]{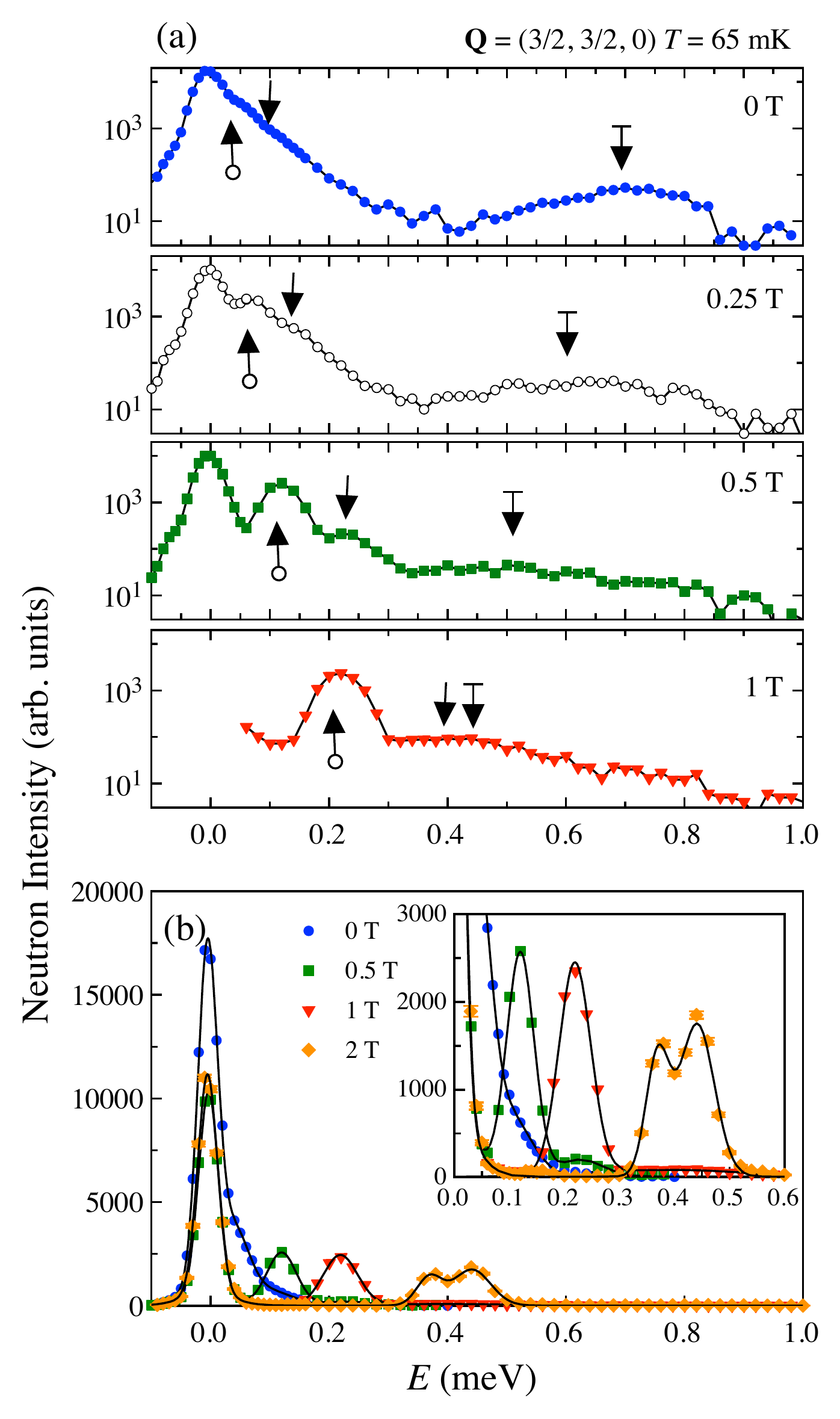}
\caption{Energy spectra at ${\bf Q}=(3/2, 3/2, 0)$ at $T=65$~mK  and various magnetic fields ${\bf H}\parallel[001]$ with the intensity 
on a log scale (a) and on a linear scale (b). The inset shows a magnified view on the 0--0.6 meV energy range. 
Lines in panel (b) are  fits described in the text.}
\label{INS3}
\end{figure}

The evolution of the excitation spectrum in an applied field is shown in Fig.~\ref{INS3} for $\textbf{Q}=(3/2,3/2,0)$ at $T=65$~mK. 
Figure~\ref{INS3}(a) presents the intensity on a log scale to emphasize the presence of three inelastic modes. 
Their positions change with field as indicated by different arrows. 
The two lowest energy modes at $E_0$ and $E_1$ move to higher energy, the first one being better defined when escaping from 
the elastic line. The higher-energy mode $E_2$, initially at 0.7 meV moves to lower energy and merges with the second mode 
at around 1 T. Figure~\ref{INS3}(b) shows this evolution on a linear scale which better illustrates the behavior of the intense elastic line 
and of the lowest energy inelastic mode at $E_0$. 
The inelastic line is fully resolved at 0.5 T, but exhibits broadening with increasing field and eventually gives rise 
to a double peak structure at 2 T,  see the inset in Fig. \ref{INS3}(b). 
As for the elastic line, its intensity is reduced at 0.25 T compared to zero field and does not further change up to 2 T. 
This observation indicates that the diffuse magnetic scattering seen in the D7 data is fully suppressed at 0.25 T.

\begin{figure}
\centering
\includegraphics[width = 0.85\columnwidth]{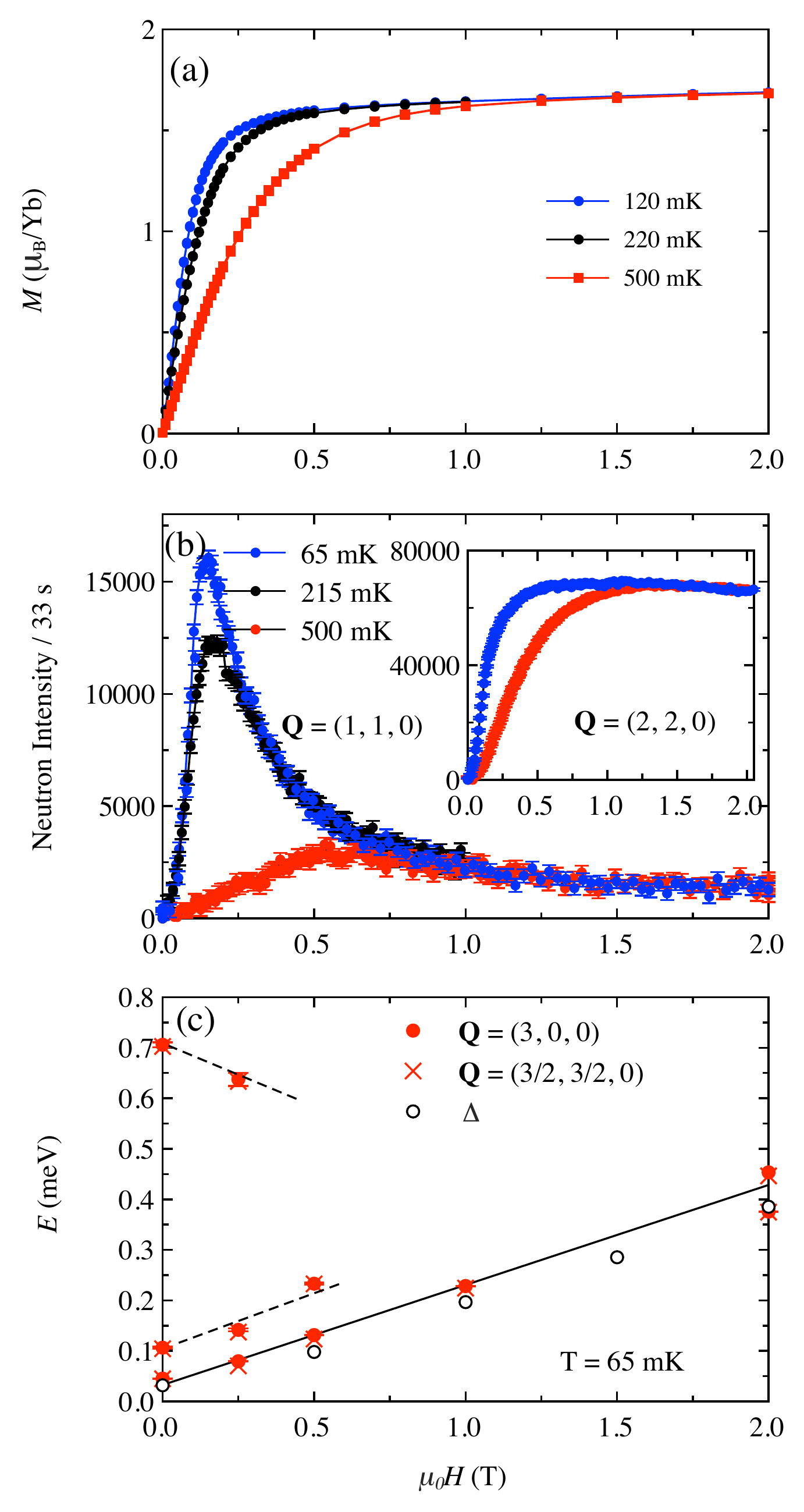}
\caption{Field dependence of physical parameters (${\bf H}\parallel [001]$):  (a) uniform magnetization;
(b) intensities of the Bragg peaks $(1, 1, 0)$  (main panel) and  (2, 2, 0)  (inset);
(c) energies of the three excitation modes obtained from the INS measurements. The full line is a linear fit, the dashed lines are guides to the eye. 
Open circles show the Schottky gap obtained from the specific heat data.}
\label{INS4}
\end{figure}

The field dependence of the peak positions is shown in Fig.~\ref{INS4}(c). 
The energy of the first mode, which is the only one well-defined in the whole field range, is fitted with a linear variation. 
The intercept at 0 T is 0.032(5)~meV and the slope is 0.198(5) meV/T \cite{Note1}.
It is to be noted that the energy of the first mode determined in this way is found to be lower than the one obtained 
by fitting the raw data in Fig. \ref{INS1}, {\it i.e.} 0.032(5) versus 0.045(1) meV. 
The slope conversion in terms of magnetic moments gives a value of 3.41 $\mu_{\rm B}$. This suggests that the $E_0$ field dependence is associated 
with the Zeeman splitting $2\mu H$, giving a moment $\mu$ of $1.70~\mu_{\rm B}$ consistent with the magnetization measurements 
below the saturation (between 2 and 8 T, the magnetization evolves from 1.70 to 1.76~$\mu_{\rm B}$). 

\begin{figure}
\centering
\includegraphics[width = 0.95\columnwidth]{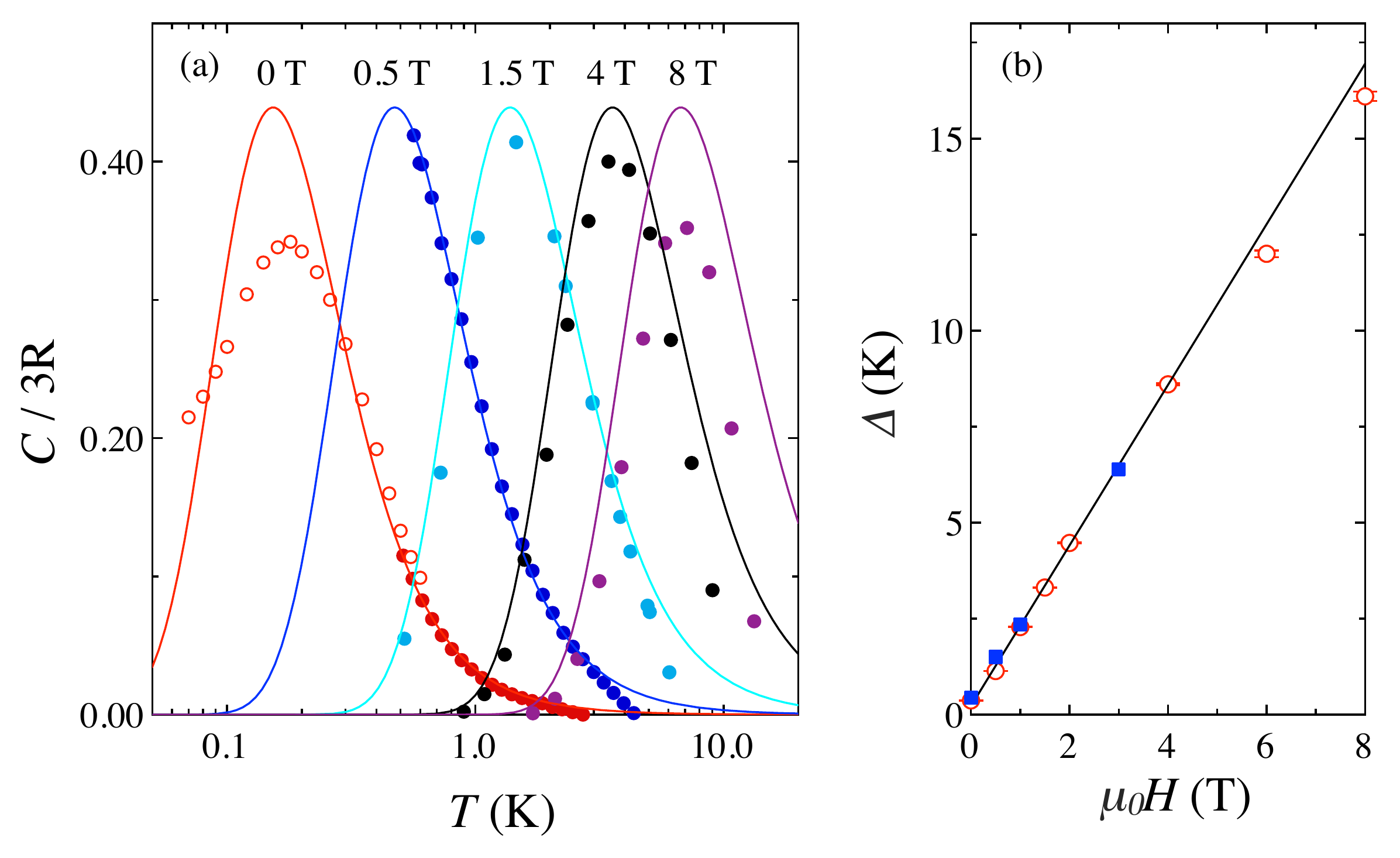}
\caption{(a) Temperature dependence of the specific heat of a ceramic sample for several magnetic fields. 
The open circles are the zero-field data by Filippi {\it et al.}\ \cite{Filippi80}. Lines are fits 
to the Schottky anomaly function. (b) Field dependence of the Schottky gap $\Delta$, open circles;
solid squares are the additional single crystal data obtained for ${\bf H}\parallel [001]$. The solid line is a linear fit 
to the low-field  data up to 4 T.}
\label{Spe}
\end{figure}

A further evidence of the Zeeman splitting effect is found in specific heat measurements, see Fig.~\ref{Spe}. 
With increasing field, the broad peak in $C(T)$ shifts to higher temperatures. The data can be fitted to a 
Schottky anomaly and the field dependence of the corresponding gap is shown in Fig.~\ref{Spe}(b). 
A linear variation can be traced up to $\mu_0H =4$~T. A similar field variation of the Schottky anomaly 
is also found for a single crystal sample (solid squares in the figure). The obtained gap is compared to the lowest energy excitation 
in Fig.~\ref{INS4}(c) and a good matching is obtained.
Indeed, as seen above, under magnetic field (already at 0.25 T), the diffuse magnetic elastic scattering 
disappears so that the low energy correlations existing in zero field are suppressed. It is thus natural that 
the specific heat Schottky anomaly reflects the Zeeman splitting of the INS spectrum. 
At higher fields (6 and 8~T), the fit to a Schottky anomaly is less accurate: the specific heat peak is narrower and 
of smaller intensity. This hints for a more complex picture than a simple two level system description.

In order to shed light on the field induced behavior of Yb$_{3}$Ga$_{5}$O$_{12}$, neutron diffraction measurements were 
conducted on D23.
The lattice parameter obtained from the orientation matrix of the single crystal sample is $a=12.19$~\AA.
The focus was on the magnetic field variations of the $(2,2,0)$ and $(1,1,0)$ Bragg reflection intensities. 
The inset of Fig. \ref{INS4}(b) shows the field variation at 65 and 500~mK of the (2, 2, 0) peak intensity. This peak, allowed by symmetry for the 24c Yb site, probes the ferromagnetic component of the system. Indeed, its field dependence strongly resembles the low temperature magnetization shown in Fig.~\ref{INS4}(a) in the low field region. At 65~mK, the signal increases up to about 0.5~T and then slowly decreases.
The field dependence of the $(1, 1, 0)$ peak is shown in Fig. \ref{INS4}(b). (1, 1, 0) is a forbidden peak for the 24c position and a magnetic field induced signal thus points to a breaking of the usual uniform field induced symmetry in the achieved spin configuration. At low temperature, the amplitude of the $(1, 1, 0)$ peak increases fast and reaches a maximum at 0.15 T. This intermediate state corresponds to a canting of some magnetic moments. The intensity decreases above 0.15 T, but remains finite and almost stable above 1 T, which matches the slow decrease of the (2,2,0) peak at high field. The magnetization itself never saturates up to 8~T (not shown), a small but finite slope being observed above 1 T. In the following, the phase at 2 T is nonetheless named saturated for convenience.

\begin{figure}
\centering
\includegraphics[width = 0.99\columnwidth]{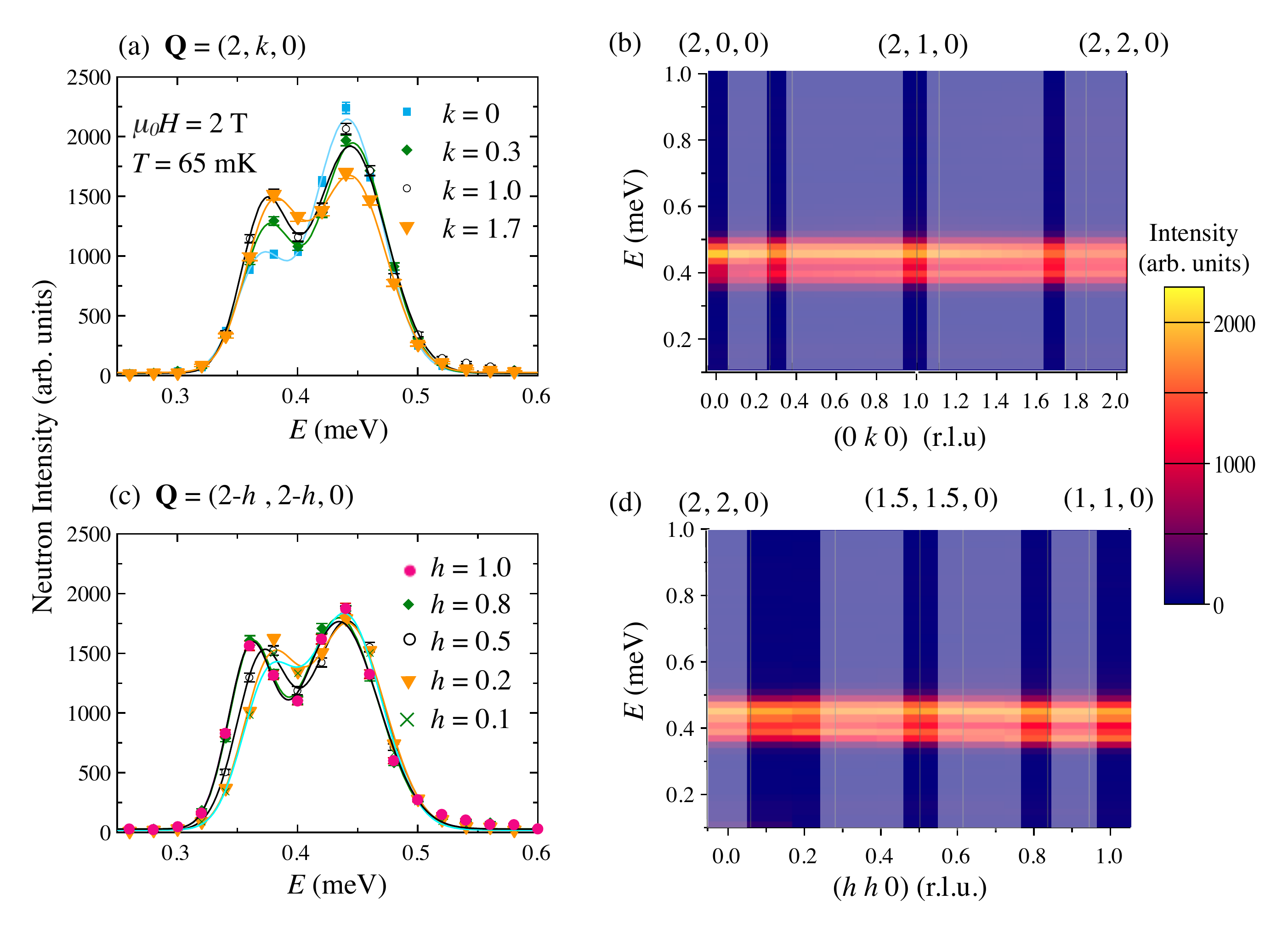}
\caption{Inelastic neutron scattering measurements at $T=60$~mK and $\mu_0 H =2$~T: (a) and (c)
the energy spectra along ${\bf Q}=(2,\ k,\ 0)$ and ${\bf Q}=(2-h,\ 2-h,\ 0)$, respectively;
(b) and (d) show the corresponding color-coded intensity maps
as functions of $\textbf{Q}$ and energy transfer. 
In (a) and (c) the lines are Gaussian fits. In the color plots (b) and (d), the shaded regions are obtained by interpolation of the data (the grid is composed of units of size 0.1 in r.l.u. and 0.02 in meV). }
\label{INS5}
\end{figure}

The measured dispersion of low-lying magnetic excitations in the 2~T saturated phase 
is illustrated in Fig.~\ref{INS5}. 
The panels show  the energy spectra for the (0 1 0) direction  \cite{Note2}, Fig.~\ref{INS5}(a),
 and  the (1 1 0) direction, Fig.~\ref{INS5}(c),  measured  at $T=65$~mK.
They exhibit a double peak structure with an individual peak width of 0.05~meV (FWHM). 
Figures~\ref{INS5}(b) and (d) present the combined color-coded intensity plots for the two directions.
The excitation peaks are essentially dispersionless. The magnon band is centered near 0.4 meV and spans 
an energy region of 0.1~meV in width.

\section{Theory}

In magnetic materials with transition temperatures in the sub-Kelvin range the dipole-dipole interactions play a significant role. Here we give a brief theoretical account of the corresponding effects for ytterbium gallium garnet, for a general discussion see \cite{DeBell00,Majlis}. 
The Kramers ions Yb$^{3+}$ in the garnet structure possess effective $S=1/2$ moments with weakly anisotropic $g$ tensors. Its principal values in the local coordinate frame were obtained in the EPR experiments on dilute (Y$_{1-x}$Yb$_x$)$_3$Ga$_5$O$_{12}$ crystals as $g_x = 2.85$, $g_y=3.60$, and $g_z = 3.74$ \cite{Carson60}. 
For simplicity, we adopt  an angular averaged 
\begin{equation}
g =\sqrt{(g_x^2+g_y^2+g_z^2)/3} \approx 3.42 \ ,
\end{equation}  
which agrees with the low-temperature magnetization data in Sec.~IIC and in \cite{Filippi80}. 
The garnet structure can be conveniently 
represented as a bcc lattice with the primitive unit cell containing 12 magnetic ions residing at 
\begin{equation}
\bm{\rho}_1 = \Bigl(0,\frac{a}{4},\frac{a}{8}\Bigr), \ \ \bm{\rho}_2 = \Bigl(0,\frac{a}{4},\frac{5a}{8}\Bigr), 
\ \ \bm{\rho}_{3,4} = - \bm{\rho}_{1,2}
\end{equation}  
and other eight sites obtained from the above four by cyclic permutations of their coordinates. The primitive
bcc translations are defined as  ${\bf a}_1 = (-a/2,a/2,a/2)$, ${\bf a}_2 = (a/2,-a/2,a/2)$,
${\bf a}_3 = (a/2,a/2,-a/2)$ with the unit cell volume $V_{\rm uc} = a^3/2$. 
The lattice constant  value $a= 12.19$ \AA\ (Sec.~\ref{mag_exp}) is used in the following.

\subsection{Dipolar interactions and the Curie-Weiss law} 

The starting point is a standard relation for the magnetic susceptibility tensor
\begin{equation}
\chi^{\alpha\beta} = \frac{(g  \mu_B)^2}{k_B T} \sum_{i,j} \langle S^\alpha_i S^\beta_j\rangle \ ,
\end{equation}  
expressed via zero-field averages of the angular momentum operators (effective spins),
$k_B$ is the Boltzmann constant. We need to consider only the diagonal elements $\chi^{\alpha\alpha}$,
which determine the longitudinal magnetization. 

At high temperatures the magnetic susceptibility can be expanded in powers of the inverse temperature
\begin{equation}
\chi = \frac{C_1}{T} + \frac{C_2}{T^2} +  \ldots \simeq \frac{C_1}{T-\theta_{CW}} 
\end{equation} 
with the Curie constant $C_1 = (g \mu_B)^2S(S+1)N/3k_B$ and the Curie-Weiss temperature  $\theta_{CW} = C_2/C_1$. 
The second-order coefficient $C_2$ in the high-temperature series expansion is expressed as
\begin{equation}
C_2 =  -\frac{(g\mu_B)^2}{k_B^2}\sum_{i,j}\Bigl[ \langle S^\alpha_i S^\alpha_j\hat{\cal H}\rangle_0 - \langle S^\alpha_i S^\alpha_j\rangle_0 \langle \hat{\cal H}\rangle_0\Bigr] \ ,
\label{C2}
\end{equation} 
where $\hat{\cal H}$ is the magnetic Hamiltonian in the absence of an external field and $\langle\ldots\rangle_0$ denotes paramagnetic averaging. As Eq.~(\ref{C2}) implies, different magnetic interactions contribute additively into $\theta_{CW}$. Thus, it is possible to separate the dipole shift from other effects.

The dipole-dipole Hamiltonian is given by
\begin{equation}
\hat{\cal H}_{\rm dip} = \frac{1}{2}(g\mu_B)^2 \sum_{i,j} \biggl[\frac{{\bf S}_i\cdot{\bf S}_j}{|{\bf r}_{ij}|^3}  - 
\frac{3({\bf S}_i\cdot{\bf r}_{ij})({\bf S}_j\cdot{\bf r}_{ij})}{|{\bf r}_{ij}|^5}\biggr] .
\label{Hdip}
\end{equation}
Substituting $\hat{\cal H}_{\rm dip}$ into Eq.~(\ref{C2}) we obtain 
the Curie-Weiss temperature for the  component $\chi^{\alpha\alpha}$  as
\begin{equation}
\theta^{\alpha\alpha}_{CW} = -\frac{(g\mu_B)^2}{3Nk_{\rm B}}S(S+1) \sum_{i,j}  \biggl[ \frac{1}{|{\bf r}_{ij}|^3} - 
\frac{3(r^{\alpha}_{ij})^2}
{|{\bf r}_{ij}|^5}\biggr].
\label{CW}
\end{equation}
Here $N$ is the number of magnetic ions and the sum over both $i$ and $j$ is kept since the result of a partial summation $\sum_j$ may vary for magnetic ions {\it inside} the unit cell. The above expression agrees with the previous works \cite{Daniels53,Gingras00}.
We further separate dimensional and  dimensionless parts in Eq.~(\ref{CW}) as
\begin{eqnarray}
\theta^{\rm dip}_{CW} & = & - \frac{J_d}{3k_B}\, S(S+1)\, \frac{a^3}{N} \sum_{i,j}  \frac{|{\bf r}_{ij}|^2 - 3z_{ij}^2}
{|{\bf r}_{ij}|^5} ,
\label{CWd} \\
&& \frac{J_d}{k_B} =  \frac{(g\mu_B)^2}{a^3}\approx 4.02~\textrm{mK} \ .
\nonumber
\end{eqnarray}
The strength of dipolar coupling between adjacent ions is estimated as
$E_{nn}= (8/\sqrt{6})^3 J_dS^2 / k_B = 35$~mK.

Lattice sums in Eqs.~(\ref{CW}) and (\ref{CWd}) are conditionally convergent, {\it i.e.}, the outcome depends on the shape of a chosen region.  
The standard way to treat this problem is to use the Lorentz sphere construction,  which explicitly separates the shape-dependent 
demagnetization part \cite{Majlis}.
We omit the demagnetization contribution assuming that a corresponding correction is included in the  experimental data. 
The rest is represented as a discrete-lattice sum inside the Lorentz sphere and its surface integral. 
In cubic crystals the lattice sum vanishes [this applies only to the double sum in Eq.~(\ref{CWd})] and one is left with the surface integral:
\begin{equation}
 \frac{1}{N} \sum_{i,j}   \frac{|{\bf r}_{ij}|^2 - 3z_{ij}^2}
{|{\bf r}_{ij}|^5}  = -\frac{4\pi}{3v_0} \ ,
\end{equation}
where $v_0= a^3/24$ is the volume per one magnetic ion in a garnet lattice.  
Thus, we obtain for Yb$_3$Ga$_5$O$_{12}$
\begin{equation}
\theta_{CW}^{\rm dip} = 32\pi J_d\, \frac{S(S+1)}{3k_B} \approx 101\ \textrm{mK} \ .
\label{CWdd}
\end{equation}
This theoretical estimate is very close to the experimental value $\theta_{CW}^{\rm exp} =  97$~mK, Sec.~\ref{mag_charac}. 
The remaining small difference can be attributed to an uncertainty of our experimental measurements, to the Van Vleck contribution, 
or an imprecise knowledge of the $g$-factor. Hence, we conclude that magnetic properties of  Yb$_3$Ga$_5$O$_{12}$ can be well 
approximated by the dipolar spin-1/2 model with nearly isotropic  
$g\sim 3.42$. A possible exchange coupling between the nearest-neighbor ytterbium ions does not exceed 1--5~\%\ of the dipole-dipole 
interaction between them.

\subsection{Magnons in the saturated state}

In a strong field, magnetic moments become almost parallel to the field direction. Since we choose an
isotropic $g$-tensor for Yb$^{3+}$ ions, the only source of a remaining small canting is the dipolar interaction.
We neglect this effect, which is justified once the Zeeman gap $\Delta_H$ in the magnon spectrum is sufficiently large 
in comparison with $J_d$ ($E_{nn}$).

Assuming that the external field $H$  and the local moments are oriented along the cubic $\hat{\bf z}$ axis we use 
the Holstein-Primakoff transformation to transform  the Hamiltonian ({\ref{Hdip}) to the quadratic bosonic form.
The position of a magnetic ion is expressed as ${\bf r}_{ni} = \bm{\rho}_{n} + {\bf r}_{i}$, where  $\bm{\rho}_{n}$ is 
its location in the primitive unit cell and ${\bf r}_{i}$ is the cell position.
Then,
\begin{eqnarray}
\hat{\cal H}_2 & =  & \frac{1}{2}\, J_dS\sum_{ni,mj} \Bigl[-D^{zz}_{ni,mj}(a_{ni}^\dagger a^{_{}}_{ni} +
a_{mj}^\dagger a^{_{}}_{mj}) 
\nonumber \\
& + &\frac{1}{2} (D^{xx}_{ni,mj} + D^{yy}_{ni,mj})(a_{ni}^\dagger a^{_{}}_{mj} +a_{mj}^\dagger a^{_{}}_{ni}) \Bigr] 
\nonumber \\
& + & H\sum_i a_{ni}^\dagger a^{_{}}_{ni}  \ .
\label{H2}
\end{eqnarray}
where the dipolar matrix is defined as
\begin{equation}
D^{\alpha\beta}_{ni,mj}  = \frac{\delta^{\alpha\beta}}{|{\bf r}_{ij}^{nm}|^3} - 
\frac{3(r_{ij}^{nm})^\alpha (r_{ij}^{nm})^\beta}{|{\bf r}_{ij}^{nm}|^5}
\end{equation}
with ${\bf r}_{ij}^{nm} = {\bf r}_{ni} - {\bf r}_{mj}$ and all distances measured in  units
of the lattice constant $a$.
We have excluded  from Eq.~(\ref{H2})  the anomalous terms $a_{ni} a_{mj}$ ($a^\dagger_{ni} a^\dagger_{mj})$. These as well as the linear terms  contribute quadratically to the magnon energy $\varepsilon_{\bf k}$: $\sim J_d^2/\Delta_H$ and can be neglected in comparison to the linear in $J_d$ dispersion originating from the normal terms.

After the Fourier transformation, the quadratic form $\hat{\cal H}_2$ changes to
\begin{eqnarray}
\hat{\cal H}_2 & =  & \sum_{n,{\bf k}} a_{n{\bf k}}^\dagger a_{n\bf k}\bigl[H-  J_dS\sum_m D^{zz}_{nm}(0)\bigr] 
 \label{H2k} \\
& + &\frac{1}{2}\,J_dS \sum_{n,m,{\bf k}}  a_{n{\bf k}}^\dagger a_{m\bf k} \bigl[D^{xx}_{nm}({\bf k})  + D^{yy}_{nm}({\bf k}) \bigr] \ ,
\nonumber
\end{eqnarray}
where the Fourier transform of the dipolar matrix is 
\begin{equation}
D_{nm}^{\alpha \beta}(\mathbf{k}) = \sum_j D^{\alpha\beta}_{ni,mj}\; e^{-i \mathbf{k} \cdot ({\bf r}_j - {\bf r}_i)} \ .
\label{Dnm}
\end{equation}
The matrix elements $D_{nm}^{\alpha \beta}(\mathbf{k})$ are computed using the Ewald summation technique. 
The details of this method are described, for example,
in \cite{DeBell00}. Below, we give only the final expression in a general form suitable for any non-Bravais lattice:
 \begin{widetext}
\begin{eqnarray}
D_{nm}^{\alpha \beta}({\bf k}) &  = & -\frac{4k^3_c}{3\sqrt{\pi}}\,\delta_{\alpha \beta}\delta_{nm}  + \frac{4 \pi}{V_{\rm uc}}
\sum_{\mathbf{G}}{}^\prime\; \frac{(k^\alpha + G^\alpha)(k^\beta + G^\beta)}{|\mathbf{k} + \mathbf{G}|^2}\: 
e^{-|\mathbf{k} + \mathbf{G}|^2/4k_c^2}\:  e^{i(\mathbf{k} + \mathbf{G})\cdot(\bm{\rho}_m - \bm{\rho}_n)}  
 \label{Ewald} \\  
& + & \sum_{\bf R} e^{-i\mathbf{k}\cdot{\bf R}} \bigg[ \frac{\textrm{erfc}(k_c R_{nm})}{R_{nm}^3}
\Big( \delta_{\alpha \beta} -  \frac{3 R_{nm}^\alpha R_{nm}^\beta}{R_{nm}^2} \Big) - 
\frac{2 k_c}{\sqrt{\pi}  R_{nm}^2}\,e^{-k^2_c R_{nm}^2}
\Big(2 k_c^2 R_{nm}^\alpha R_{nm}^\beta -\delta_{\alpha \beta} +  \frac{3 R_{nm}^\alpha R_{nm}^\beta}{R_{nm}^2}\Big)\bigg].
\nonumber  
\end{eqnarray} 
\end{widetext}
Here, $\textrm{erfc}(x)$ is the complementary error function, the first sum is performed over the reciprocal lattice points
${\bf G} = m_1 {\bf b}_1 + m_2 {\bf b}_2 + m_3 {\bf b}_3$, whereas the second sum is taken over the direct lattice
${\bf R} = n_1 {\bf a}_1 + n_2 {\bf a}_2 + n_3 {\bf a}_3$,  ${\bf R}_{nm} = {\bf R}  + \bm{\rho}_n - \bm{\rho}_m$, and $k_c$ is an arbitrary  cutoff parameter. Independence of a final result on the actual value of $k_c\in [0.2,2]$ can serve as a convenient consistency check. For arbitrary ${\bf k}\neq {\bf G}$, both sums in Eq.~(\ref{Ewald}) rapidly converge. However, the term with ${\bf G}=0$ in the first sum becomes ill-determined for ${\bf k}\to 0$ and has to be dropped out, which is indicated by a prime near the summation symbol. 
This regularization procedure is equivalent to neglecting the demagnetization effects, {\it i.e.}, assuming that a  magnetic sample has cylindrical or slab geometry with a parallelly applied field.

\begin{figure}[ht!]
\begin{center}
\includegraphics[width = 0.8\columnwidth]{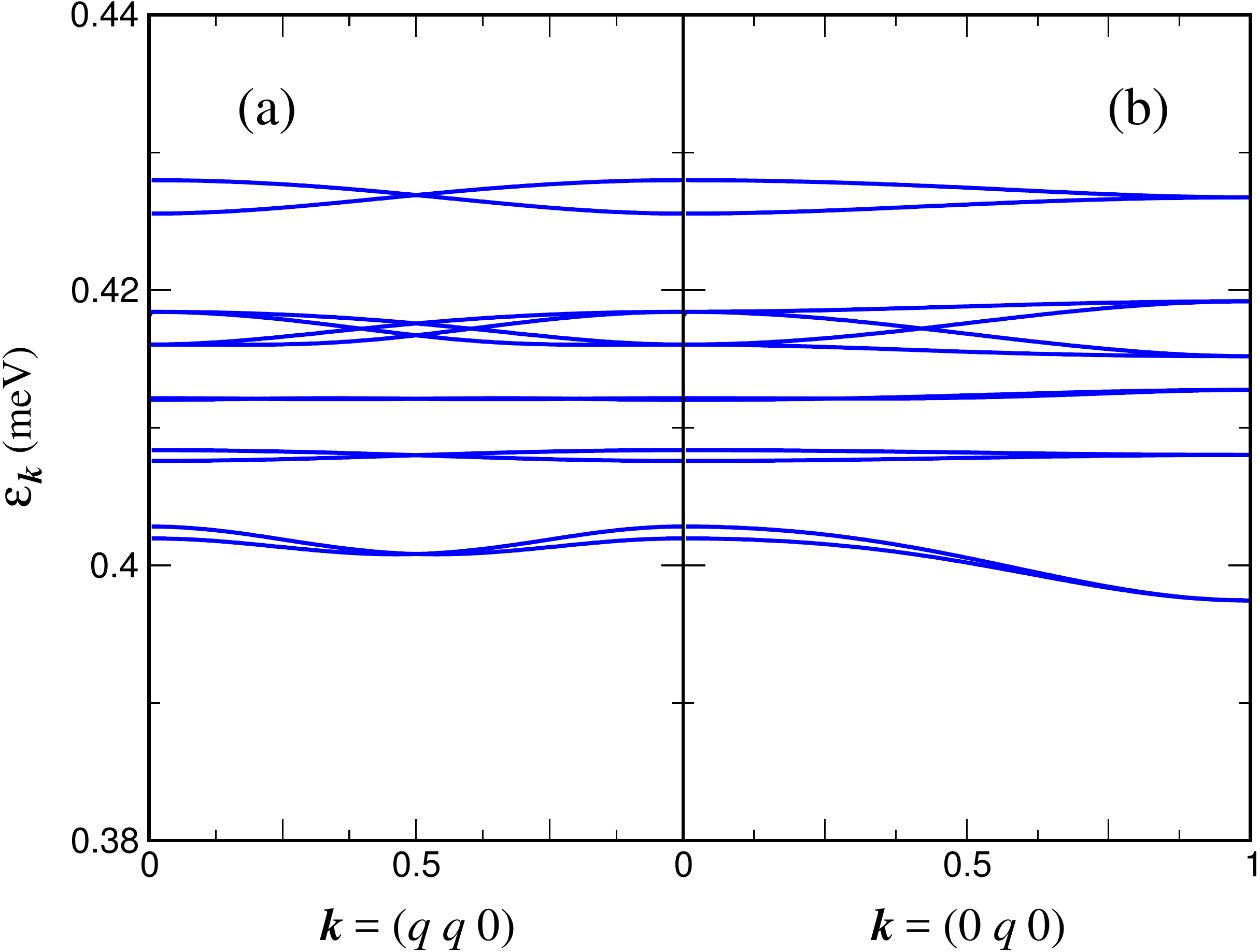} 
\\[4mm]
\includegraphics[width = 0.8\columnwidth]{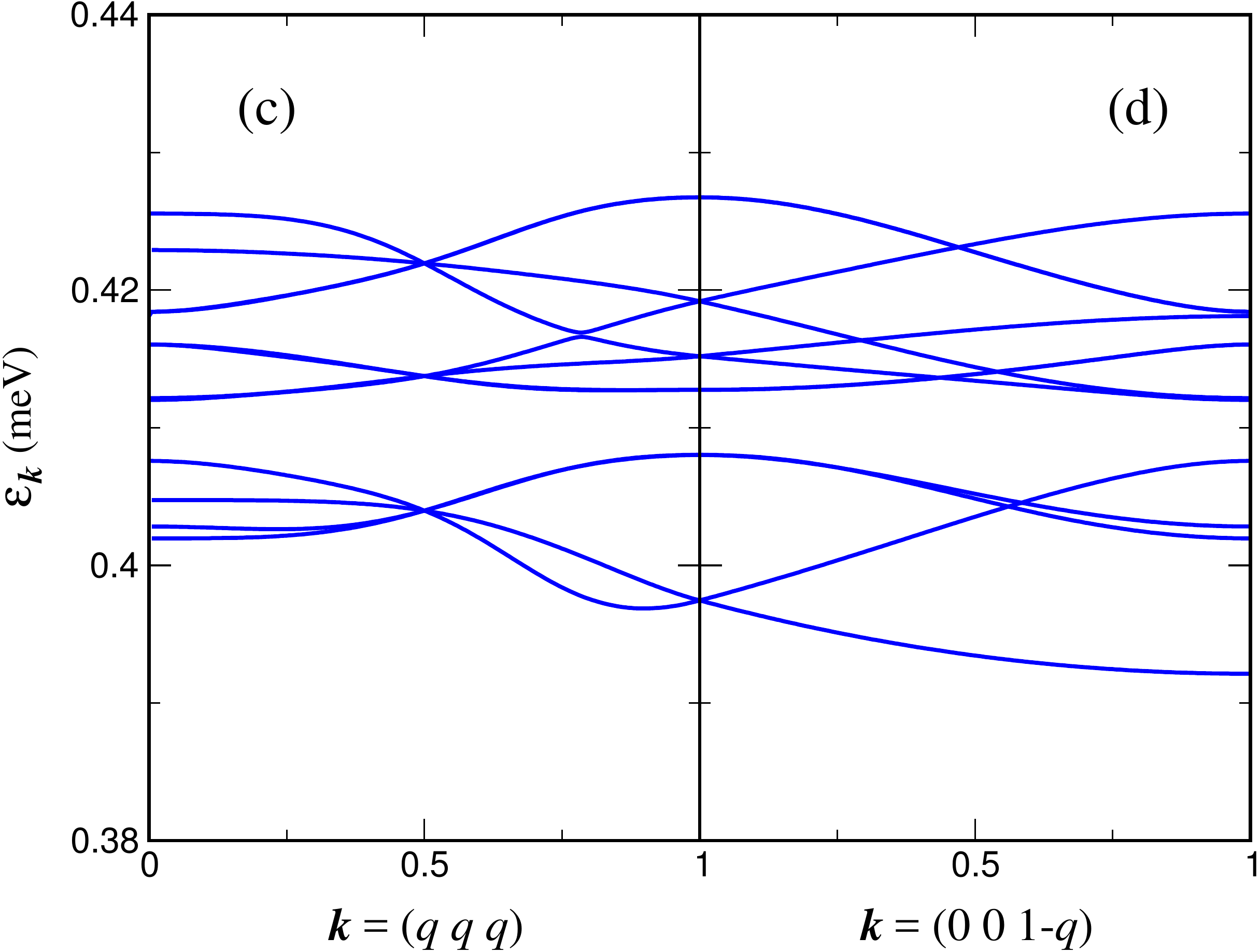}
\end{center}
\caption{Magnon spectra along the symmetry lines in the Brillouin zone computed for the dipolar spin model 
with ${\bf H}\parallel [001]$, $\mu_0H=2$~T. The wavevector components are measured in units of $2\pi/a$.}
\label{fig:disp}
\end{figure}

The magnon spectra  were computed by numerical diagonalization of  Eq.~(\ref{H2k}).
Figure~\ref{fig:disp} displays results for an applied field of $\mu_0H = 2$~T with ${\bf H}\parallel[001]$
by showing the magnon energy $\varepsilon_{\bf k}$ for the symmetry directions in the Brillouin zone.
Since the primitive unit cell contains 12 magnetic sites, the excitation spectra have twelve branches 
with weak dispersion proportional to $J_d$. The branches are gathered
into several groups depending on the momentum direction. The total magnon bandwidth $\sim 0.04$~meV is an order of 
magnitude smaller than the gap $\Delta_H\sim 0.4$~meV. Such a large separation of the energy scales justifies 
the approximation adopted above. However, for weak fields $\mu_0 H\alt 0.4$~T the neglected
anomalous terms as well as the local spin canting have to be taken into account again.

An interesting feature of the computed  spectrum is multiple crossing points
between different magnon branches.  Some of them correspond to crossing of three or even more 
branches, {\it e.g.}, at $(1/2,1/2,1/2)$ in Fig.~\ref{fig:disp}(c).
Such conical crossing points may be stable in the cubic garnet crystals,
if the Bloch states in the meeting branches transform according to a three-dimensional irreducible representation.
As a result, magnon bands may possess interesting topological properties. 
These questions deserve a separate theoretical investigation.
Comparing with experiment, we note that 
the average position of magnon bands at $\sim 0.4$~meV matches well the neutron data of Sec.~\ref{mag_exp}.
The limited instrumental resolution as well as thermal fluctuations ($T \sim  2 E_{nn}$) cause
a smearing effect producing a broad double-peak structure in the experimental data of Fig.~\ref{INS5}
instead of individual magnon lines. This structure is consistent with the observed tendency of magnon
branches to further bunch into several groups.
Still, the experimental bandwidth is twice larger than the calculated one. Additional future measurements and calculations 
may be needed to clarify the origin of the observed discrepancy.

\section{Discussion}

Previously, the diffuse scattering was measured in the (1 -1 0)--(1 1 2) plane and analyzed using reverse Monte Carlo calculations 
performed for the powder average response \cite{Sandberg20}.
In our work, the diffuse scattering pattern shown in Fig. \ref{Elast1} is seen projected from a 4-fold axis.
Three peaks are found at 0.36, 1.1 and 1.95~{\AA}$^{-1}$. The first and last values coincide approximately with the ones found in Ref. \onlinecite{Sandberg20}, which reports another peak at 1.86~{\AA}$^{-1}$.
These typical ranges of Q are also characteristic of the short range correlations found in Gd$_{3}$Ga$_{5}$O$_{12}$.
This highlights the common spin configurations involved, with contributions from trimers and 10-spin loops \cite{Deen15}.
Furthermore, we show that the corresponding correlations in Yb$_{3}$Ga$_{5}$O$_{12}$ are significantly increased below 200~mK while a $\bf{Q}$ independent  signal (except for magnetic form factor dependence) is still present at 5 K. The temperature dependence of the elastic signal finds echo in the increase of the relaxation rate found by $\mu$SR in the range $T_{\lambda}$ to 0.4 K \cite{Dalmas03}.

The three modes of the magnetic excitation spectrum of Yb$_{3}$Ga$_{5}$O$_{12}$ we observe are in agreement with results of Sandberg {\it et al.} \cite{Sandberg20} who report rather similar energies (0.06, 0.12 and 0.7 meV).
Our field dependence analysis could however show that the first mode has a lower energy $E_0=0.032$ meV. 
A precise determination at zero field would need experiments with higher resolution. It is to be noted that the elastic and inelastic responses seem not to be correlated since the strong $\bf{Q}$-dependence of the former is not reflected in the latter. 
Interestingly, this magnetic excitation spectrum is very similar to the one observed in Gd$_{3}$Ga$_{5}$O$_{12}$ where the sequence in energies is 0.04, 0.12 and 0.58 meV \cite{Deen10}, while the two systems are expected to be different since the Yb$_{3}$Ga$_{5}$O$_{12}$ ground state is supposed to be ordered. This confirms the similar nature of the correlated state (before ordering) in both systems. 

The intermediate spin configuration we observe under field through the maximum intensity of the  $(1, 1, 0)$ Bragg peak at 0.15 T presumably corresponding to a spin canting also reminds the complicated magnetic structure observed in Gd$_{3}$Ga$_{5}$O$_{12}$ \cite{Petrenko, Deen15}. The studies in this latter compound were nevertheless hardly conclusive due to the absence of large isotopic single crystals suitable for neutron scattering experiments. We therefore expect that further studies in Yb$_{3}$Ga$_{5}$O$_{12}$ will help elucidate the complex field induced magnetic structure.
In contrast to Gd$_{3}$Ga$_{5}$O$_{12}$, we do not observe any signatures of field induced transitions in magnetization measurements so that the anomalous field variation of the Bragg peaks could be indicative of a crossover regime. In addition, our ac susceptibility data do not show evidence for a magnetic freezing in Yb$_{3}$Ga$_{5}$O$_{12}$ down to 70 mK contrary to the Gd$_{3}$Ga$_{5}$O$_{12}$ case \cite{Schiffer95}.

The most original contribution of the present study is the focus on the spin dynamics under magnetic field.
The experimental ferromagnetic Curie-Weiss temperature is quantitatively well accounted for the dipolar interactions.
The magnetic excitation spectrum in the saturated phase is also well reproduced with a magnon calculation based on a dipole-dipole Hamiltonian.
A challenge is the understanding of the spin excitation at zero field and the global magnetic behavior under field below the saturation. An interesting question is to know if the dipolar interactions can explain the canting away from the applied field below the saturated phase as proposed for Gd$_{3}$Ga$_{5}$O$_{12}$ (See Fig. 1(c) of Ref. \onlinecite{Ambrumenil}).

Finally, we have demonstrated that the field dependence of the specific heat  
is governed by the field evolution of the band of low-lying magnon excitations.  
This highlights presence of an additional microscopic contribution to the enhanced magnetocaloric effect 
in Yb$_{3}$Ga$_{5}$O$_{12}$  determined by collective excitation modes, in line with the theoretical
proposal \cite{Zhito03}.
Indeed, in paramagnet salts that are widely used for ADR applications
the  excitation spectrum consists solely of a quasielastic component with width proportional to $k_BT$.
Within resolution of the neutron experiments it appears as an elastic signal  \cite{Boothroyd} and the spectral weight 
is simply transferred to a ferromagnetic Bragg peak once the applied field exceeds $k_{B}T$.

\section{Conclusion}

Our study demonstrates that the magnetic response of Yb$_{3}$Ga$_{5}$O$_{12}$  is composed of two distinct contributions: 
from the ${\bf Q}$-dependent elastic signal and  from almost dispersionless inelastic modes. This landscape appears to be quite 
similar to the one found in another garnet, namely Gd$_{3}$Ga$_{5}$O$_{12}$, extensively studied in the past.
The present work reveals the dominant role of dipolar interactions in Yb$_{3}$Ga$_{5}$O$_{12}$.
The theoretical calculations for the spin-1/2 dipolar model reproduce the gross features of the experimental 
magnon spectrum in the saturated phase and quantitatively account for 
the Curie-Weiss temperature. Thus, ytterbium gallium garnet provides an interesting example of a quantum
dipolar magnet on the hyperkagome lattice. Further experimental and
theoretical investigations of the spin dynamics are necessary to elucidate the interplay between 
its highly symmetric frustrated geometry and the long-range dipolar interactions.
In addition, the obtained results confirm that strong spin correlations in a frustrated geometry
can enhance the magnetocaloric response. We hope that this observation will motivate a continuing search of suitable 
frustrated magnetic materials for ADR applications.

\section{Acknowledgements}

The INS data collected at the ILL for the present work are available at Ref. \onlinecite{DOI1,DOI2}. 
The sample orientation for the neutron experiments was made on IN3 and Orient Express.
E.L., J.-P.B., C.M. and M.E.Z. acknowledge financial support from ANR, France, Grant No. ANR-18-CE05-0023.  J.-P.B., C.M. and M.E.Z. acknowledge financial support from the Materials transverse program of CEA ``CeramOx".

\appendix
\section{Fit of the Neutron data}
\label{appendix}
The elastic line does not evolve between 0.25 and 2 T and therefore all the magnetic elastic scattering is suppressed at 0.25 T.
The remaining incoherent scattering is then estimated from the 2 T data and is described by two gaussians of width 0.039 meV and 0.096 meV, the second one having an intensity of 1/18 of the first one (This reflects globally the non-gaussian profiles of the monochromator and analyser transfer functions). In all the fits of the elastic line, only an overall intensity (with the second gaussian intensity fixed at 1/18 of the first one) and an offset are free parameters. The energies of the modes given in the text are corrected for the offset. For the inelastic modes under magnetic field, the intensity and peak positions are always free parameters. The width of the lowest energy mode is fixed at 0 T to the value found at 0.25 T, where the peak is resolved and the width is then a free parameter for all finite fields. For the second mode, the width is fixed at 0, 0.25 and 0.5 T and free at 1 T when it merges with the third mode. For the third mode, all parameters are free.

\end{document}